# Continuous Architecting with Microservices and DevOps: A Systematic Mapping Study

Davide Taibi[1], Valentina Lenarduzzi[1], and Claus Pahl[2]

[1] Tampere University, Tampere, Finland
{davide.taibi,valentina.lenarduzzi}@tuni.fi
[2] Free University of Bozen-Bolzano, Bolzano, Italy
claus.pahl@unibz.it

**Abstract.** Context: Several companies are migrating their information systems into the Cloud. Microservices and DevOps are two of the most common adopted technologies. However, there is still a lack of under- standing how to adopt a microservice-based architectural style and which tools and technique to use in a continuous architecting pipeline.

Objective: We aim at characterizing the different microservice archi- tectural style principles and patterns in order to map existing tools and techniques adopted in the context of DevOps.

Methodology: We conducted a Systematic Mapping Study identifying the goal and the research questions, the bibliographic sources, the search strings, and the selection criteria to retrieve the most relevant papers.

Results: We identified several agreed microservice architectural prin- ciples and patterns widely adopted and reported in 23 case studies, together with a summary of the advantages, disadvantages, and lessons learned for each pattern from the case studies. Finally, we mapped the existing microservices-specific techniques in order to understand how to continuously deliver value in a DevOps pipeline. We depicted the current research, reporting gaps and trends.

Conclusion: Different patterns emerge for different migration, orches- tration, storage and deployment settings. The results also show the lack of empirical work on microservices-specific techniques, especially for the release phase in DevOps.

## 1 Introduction

Software is becoming more complex and development processes are evolving to cope with the current fast-changing requirements imposed by the market, with short time-to-market and quickly evolving technologies. Continuous soft- ware engineering, and in particular DevOps, tries to address these aspects, sup- porting developers with a set of continuous delivery practices and tools to contin- uously deliver value, increasing delivery efficiency and reducing the time intervals between releases [3]. However, traditional monolithic architectures are not easily applicable to this environment and new architectural styles need to be consid- ered. In order to adopt DevOps, the architectural style adopted must be designed with an agile focus; for this purpose, the Microservices [10] architectural style is suitable for this continuous architecture setting.

Microservices are relatively small and autonomous services deployed indepen- dently, with a single and clearly defined purpose [10]. Because of their indepen- dent deployability, they have a lot of advantages for continuous delivery. They can be developed in different programming languages, they can scale indepen- dently from other services, and they can be deployed on the hardware that best suits their needs. Moreover, because of their size, they are easier to maintain and more fault-tolerant since the failure of one service will not break the whole system, which could happen in a monolithic system [12].

DevOps (Development and Operations) is a set of continuous delivery prac- tices aimed at decrease the delivery time, increasing the delivery efficiency and reducing time among releases while maintaining software quality. It combines software development, quality assurance, and operations [3]. DevOps includes a set of steps of the development process (plan, create, verify, package) and of the operational process (release, configure, monitor), combining the activities com- monly performed by the development teams, quality assurance and operations teams. In order to adopt the DevOps practices, the architectural style of the system must be design with an agile focus and the microservice architectural style is one of the most suitable architectural style to cope with them [2].

Despite both the microservice architectural style and DevOps being widely used in industry, there are still some challenges in understanding how to develop such kinds of architectures in a continuous software engineering process [2]. In this work, we extend our previous mapping study on architectural patterns for microservices [16].

The goal of this work is two-fold: First we aim to characterize the differ- ent microservice architectural styles reported in the literature both as propos- als and case studies on implementations. Then we aim to map the reported microservices-based techniques that can be applied to the DevOps pipeline in order to identify existing gaps. Therefore, we designed this work as a Systematic Mapping Study [13, 19]. A previous systematic mapping has been published by Pahl and Jamshidi [11] aimed at classifying and comparing the existing research body on microservices mainly considering non peer-reviewed content from web blogs. Our study differs in the following ways:

- *Focus*: We focus on suggested architectural style definitions, emerging pat- terns and mapping microservices development to the DevOps pipeline, while [11] focused on initially characterizing the available body of research and [16] focused only on architectural styles.
- *Comprehensiveness* : We included results from eight bibliographic sources and papers from the citations of the retrieved papers [19] to increase the paper base. Moreover, we included papers published up to 2016;

- *Systematic approach*: We conducted a Systematic Mapping Study implementing the protocol defined in [13], followed by a systematic snowballing process using all references found in the papers [19];
- *Quality Assessment* : Although this is not a Systematic Literature Review [8], we include only peer-reviewed contributions or non peer-reviewed papers only in case the number of their citations in peer-reviewed ones is higher than the average citations.

The contribution of our study can be summarise as follows:

- Classification of the existing microservice architectural styles and patterns;
- Analysis of advantages and disadvantages of different architectural style principles and patterns based on their implementations reported in the literature;
- Classification of microservice techniques for DevOps;
- Identification of research gaps and trends.

The paper is structured as follows. In Sect. 2 we describe the methodology used. Section 3 shows the results obtained. In Sect. 4 we discuss the results. Section 5 identifies threats to validity. Section 6 end with some conclusions.

## 2 Methodology

We used the protocol defined by Petersen [13] in combination with the systematic snowballing process [19].

### 2.1 Goals and Research Questions

We define our research goals as follows:

**Goal 1:** *Analyze* the architectural style proposals
*for the purpose of* comparing them and related implementations
*with respect to* their advantages and disadvantages
*in the context of* cloud-native software implementation.
**Goal 2:** *Characterize* microservices-specific techniques
*for the purpose of* mapping them to the DevOps process
*with respect to* identifying and comparing different techniques for different stages
*in the context of* cloud-native software implementation. Regarding G1, we derived the following research questions:

- RQ1: Which are the different microservices-based architectural styles?
- RQ2: What are the differences among the existing architectural styles?
- RQ3: Which advantages and disadvantages have been highlighted in implementations described in the literature for the identified architectural styles?

Regarding G2, we derived the last research question:

- RQ4: What are the different DevOps-related techniques applied in the microservices context?

## 2.2 Search Strategy

**Bibliographic Sources and Search Strings.** We identified the relevant works in eight bibliographic sources as suggested in [9]: ACM Digital Library, IEEE Xplore Digital Library, Science Direct, Scopus, Google Scholar, Citeeser library, Inspec and Springer Link. We defined the search strings based on the PICO terms of our questions [9] using only the terms Population and Intervention. We did not use the Outcome and Comparison terms so as not to reduce research efficiency of the selected search strings (Table 1). We applied the following queries adapting the syntax to each bibliographic source:
**RQ1-3**: (microservice* OR micro-service*) AND (architect* OR migrat* OR modern* OR reengineer* OR re-engineer* OR refactor* OR re-factor* OR rearchitect* OR re-architect* OR evol*).
**RQ4**: (microservice* OR micro-service*) AND (DevOps OR Develop* OR Creat* OR Cod* OR verif* OR test* OR inspect* OR pack* OR compil* OR archiv*; releas* OR configur* OR deploy* OR monitor* OR performance* OR benchmark*).

The symbol * allows to capture possible variations in search terms such as plural and verb conjugation.

**Table 1.** Search strings - PICO structure [16].

| Population | Intervention - terms |
|---|---|
| P: microservice | microservice*; micro-service* |
| I: DevOps; architecture; migration | architect*; migrat*; modern*; evol*; reengineer*; re-engineer*; refactor*; re-factor*; rearchitect*; re-architect*; DevOps; Develop*; Creat*; Cod*; verif*; test*; inspect*; pack*; compil*; archiv*; releas*; configur*; deploy*; monitor*; performance*; benchmark; |

**Inclusion and Exclusion Criteria.** We defined the selection criteria based on our RQs considering the following inclusion criteria:

*General Criteria:* We only included papers written in English. Moreover, we excluded papers that were not peer-reviewed. However, we also considered non peer-reviewed contributions if the number of citations in peer-reviewed papers was higher than average. The number of unique citations was extracted from the eight bibliographic sources removing non peer-reviewed ones. The selected works cover a maximum of two years and we can therefore not expect a high number of citations. For this reason, works with a high number of citations can be considered very relevant even if they are not peer-reviewed.

*Selection by Title and Abstract:* We removed all papers that do not contain any references to microservices or that use the term microservices for different purposes or in different domains (i.e. electronics, social science...);

*Selection by Full Papers:* We excluded papers that do not present any evidence related to our research questions or papers using microservices with- out any clear reference to the adopted architectural style, and microservices- based implementations that do not report any advantages and disadvantages of using microservices. For the first three RQs, we considered proposals of microservices-based architectural styles, implementations of microservices-based cloud systems, migrations of monolithic systems into cloud-native microservices- based systems, papers reporting advantages and disadvantages of microservices-based architectural styles. For RQ4, we considered papers on DevOps techniques applied in the context of microservices-based systems, and papers on *project planning*, *coding*, *testing*, *release*, *deployment*, *operation* and *monitoring* tech- niques applied in the context of microservices-based systems.

**Search and Selection Process.** The search was conducted in October 2017 including all the publications available until this period. Applying the searching terms we retrieved 2754 unique papers.

*Testing Inclusion and Exclusion Criteria Applicability:* Before applying the inclusion and exclusion criteria, we tested their applicability [9] to a subset of 30 papers (10 papers per author) randomly selected from the retrieved ones. For 8 of the 30 selected papers, two authors disagreed and a third author was involved in the discussion to clear the disagreements.

*Applying Inclusion and Exclusion Criteria to Title and Abstract:* We applied the refined criteria to remaining papers. Each paper was read by two authors and in case of disagreed and a third author was involved in the discussion to clear the disagreements. For seven papers we involved the third author. Out of 2754 initial papers, we included 85 by title and abstract.

*Backward and Forward Snowballing:* We performed the backward and forward snowballing [19], considering all the references presented in the 85 papers (858 references) and evaluating all the papers that reference the retrieved ones result- ing in one additional relevant paper. We applied the same process as for the retrieved papers. The new selected studies were included in the aforementioned 12 papers, in order to compose the final set of publication.

*Fulfill Reading:* After the full reading of the 97 papers performed by two of the authors, the paper identification process resulted in 40 peer-reviewed papers and 2 non peer-reviewed ones. The two works ([S1] and [S2]) added from the gray lit- erature have a dramatically high number of citations compared to the remaining works, with 18 and 25 citations, resp. (average number of citations=4.21). The related citations are reported together with the full references in the Appendix.

In case of [S2], we also attributed to the same work the citations obtained for [14], since this website was published with the same information two months later.

**Table 2.** The papers selection process [16].

| Selection process | #considered papers | #rejected papers | Validation |
|---|---|---|---|
| Paper extracted from the bibliographic sources | 2754 | | 10 random papers independently classified by three researchers |
| Sift based on title and abstract | | 2669 | Good inter-rater agreement on first sift (K-statistic test) |
| Primary papers identified | 85 | | |
| Secondary papers inclusion | 858 | 855 | Systematic snowballing [19] including all the citations reported in the 85 primary papers and sifting them based on title and abstract |
| Full papers considered for review | 88 | | Each paper has been read completely by two researchers and 858 secondary papers were identified from references |
| Sift based on full reading | | 46 | Papers rejected based on inclusion and exclusion criteria |
| **Relevant papers included** | **42** | | |

The selection process resulted in 42 accepted papers published from 2014 to 2016. Although the term microservice was introduced in 2011, no publications were found from 2011 to 2013. More than 65% of these papers were published at conferences, while another 23% were accepted at workshops. Only 7% of the papers were published as journal articles, and nearly 5% are non peer-reviewed websites (gray literature) (Table 2).

# 3 Results

We now summarize the pros and cons of microservice-based solutions based on their importance, considering the concerns mentioned most frequently in the papers as being important. We analyze the most common architectural style prin- ciples and patterns that emerged from the papers, also including their reported advantages and disadvantages. Moreover, we report on DevOps-related tech- niques applied. We first report on the principles of microservices architectural styles, as reflected by the literature, and then we extract and categorize the patterns defined in the surveyed literature.

We consider an *architectural style* as a set of *principles* and coarse-grained *patterns* that provide an abstract framework for a family of systems. An archi- tectural style consists of a set of architectural principles and patterns that are aligned with each other to make designs recognizable and design activities repeat- able: principles express architectural design intent; patterns adhere to the prin- ciples and are commonly occurring (proven) in practice.

## 3.1 General Advantages and Disadvantages of Microservices and Principles of the Architectural Style

The most common advantages of microservice architectures that are highlighted in the selected works are the following:

- *Increased Maintainability*. All the paper reported microservices-based implementations as the most important considered characteristic.
- *Write Code in Different Languages*. Underlines benefits of using different languages, inconsistent with monolithic applications [S13], [S34], [S11].
- *Flexibility*. Every team can select their own technology based on their needs [S30], [S14], [S38]
- *Reuse*. The creation of a component with shared features increase reusability by reducing maintenance effort since the shared component will be updated only once and the maintenance of the shared microservices, including all the related changes will be reflected by any connected microservices [S34], [S12].
- *Ease of Deployment*. The independent deployment ease the whole develop- ment and deployment processes since each microservice can be deployed sep- arately. Therefore, developers of one microservice do not need to recompile and re-deploy the whole system [S30]
- *Physical Isolation*. This is the key for scaling, provided by microservices archi- tectural style [S3] and [S38].
- *Self-Healing*. Previous safe microservice versions can replace failing services [S7], [S30].
- *Application Complexity*. Components application are commonly less complex and easier to manage thanks to the application decomposition into several components [S29]. Process mining could be highly beneficial in this context [18]
- *Unlimited Application Size*. Microservices has theoretically no size limitation that affect monolithic applications [S13].

These can be considered to form the *principles of the architectural style* as they are agreed advantages. On the other hand, several papers identified a set of issues and potential **disadvantages** to be consider during the development of a microservices-based application:

- *Testing Complexity*. More components and patterns of collaborations among them increase the testing complexity [S21], [S24], [S26], [S31], [S37], [S28].
- *Implementation Effort*. Paired with development complexity, implementing microservices requires more effort than implementing monolithic applications [S28], [S30], [S38].
- *Network issues*. Endpoints are connected via a network. Therefore, the net- work must be reliable [S41], [S14].
  - *Latency*. Network latency can increase the communication time between microservices [S14], [S11], [S9].
  - *Bandwidth*. Communication often depends on the network, implementa- tions must consider bandwidth for normal and high peak operation.

- *User Authorization*. The API exposed by the microservices need to be pro- tected with a shared user-authentication mechanism, which is often much more complex to implement than monolithic solutions [S14].
- *Time on the Market*. Monolithic solutions are easier and faster to develop. In the case of small applications, with a small number of users (hundreds or thousands), the monolith could be a faster and cheaper initial approach. A microservices-based solution could be considered in a second time once performance or other requirements grows [S11].
- *Continuously Deploy Small Incremental Changes*. The simplified deployment allows changing one issue at time and immediately deploy the system [S37].
- *Independent Monitoring*. A microservices architecture helps independently visualize the "health status" of every microservice in the system simplifying the identification of problems and speeding-up the resolution time [S37].
- *Automation Requirement*. A full DevOps stack is fundamental to manage the whole system and automate the whole process. Without the adoption of DevOps the system development would be much slower with microservices than with monolithic systems [S37].
- *High Independence*. Maintaining microservices as highly decoupled is critical to preserve independence and independent deployability.
- *Development Complexity*. Microservices require experienced developers and architects that design the system architecture and coordinate teams. Learning microservices require much longer than monolithic systems [S30].
- *Increased memory consumption.* If each service runs in its own virtual machine, as is the case at Netflix, then there is the overhead of M times as many virtual machine instances are created [S2].

## 3.2 Microservice-Based Architectural Patterns

In this section, we aim to answer RQ1, RQ2, and RQ3. From the selected works, three commonly used architectural patterns emerge. In this classification, we attribute to the different patterns the papers reporting the usage of a specific style and those where the patterns can be clearly deduced from the description.

We report the results in three Sections that classify the architectural patterns emerging from this review: In the next sub-sections, we identify and describe orchestration and coordination-oriented architectural patterns, patterns reflect- ing deployment strategies and storage options.

### The API-Gateway Pattern

*Concept:* Microservices can provide their functions in the form of APIs, and other services can make use of them by directly accessing them through an API. However, the creation of end-user applications based on the composition of different microservices requests a server-side aggregation mechanism. In the selected papers, the API-Gateway resulted as a common approach (Fig. 1).

*Origin:* The API-Gateway is an orchestration style that resembles more SOA principles than REST ones without including the Enterprise Service Bus (ESB).

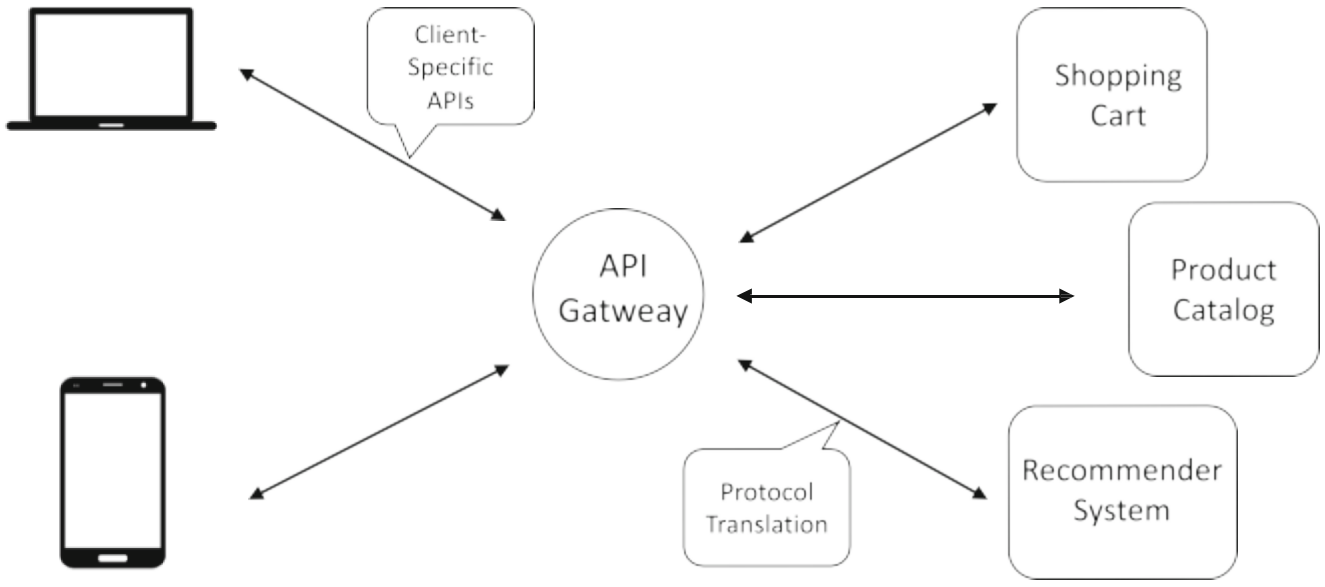

**Fig. 1.** The API-Gateway architectural pattern [16].

*Goal:* The main goal is to improve system performance and simplify interactions, therefore decreasing the number of requests per client. It acts as an entry point for the clients, carrying out their requests to the connected services, connecting the required contents, and serving them to the clients [S2].

*Properties:* The API-Gateway does not provide support for publishing, promot- ing, or administering services at any significant level. However, it is responsible for the generation of customized APIs for each platform and for optimizing com- munications between the clients and the application, encapsulating the microser- vices details. It allows microservices to evolve without influencing the clients. As an example, merging or partitioning two or more microservices only requires updating the API-Gateway to reflect the changes to any connected client. In the example depicted in Fig. 1, the API-Gateway is responsible for communicating with the different front-ends, creating a custom API for each client so that the clients can see only the features they need, which simplifies the creation of end- user applications without adding the complexity of exposing and parsing useless information.

*Evolution and Reported Usage:* The API-Gateway was named by Richardson [S2]. Ten works implemented different cloud applications based on this pattern reporting several API-Gateway specific **advantages** [S3], [S2], [S12], [S11], [S14], [S31], [S21], [S34], [S39], and [S37]:

- *Ease of Extension*. Implementing new features is easier compared to other architectures since API-Gateway can be used to provide custom APIs to the connected services. Therefore, if a services changes, only the API-Gateway needs to be updated and the connected services to the API-gateway will continue to work seamlessly [S14], [S3]
- *Market-centric Architecture*. Services can be easily modified, based on market needs, without the need to modify the whole system. [S14]
- *Backward Compatibility*. The gateway guarantees that existing clients are not hampered by interface endpoint changes on service version changes [S34].

However, **disadvantages** have also been observed for this architectural pat- tern:

- *Potential Bottleneck.* The API-Gateway layer is the single entry point for all requests. If it is not designed correctly, it could be the main bottleneck of the system [S14], [S39].
- *Implementation complexity.* The API-Gateway layer increases the complexity of the implementation since it requires implementation of several interfaces for each service [S14], [S34].
- API reused must be considered carefully. Since each client can have a custom API, we must keep track of cases where different types of clients use the same API and modify both of them accordingly in case of changes to the API interface [S34].
- *Scalability.* When the number of microservices in a system explodes, a more efficient and scalable routing mechanism to route the traffic through the ser- vices APIs, and better configuration management to dynamically configurate and apply changes to the system will be needed [S37].

## The Service Registry Pattern

*Concept:* The communication among multiple instances of the same microservice running in different containers must be dynamically defined and the clients must be able to efficiently communicate to the appropriate instance of the microser- vice. Therefore, in order to connect to an existing service, a service-discovery mechanism is needed [S2].

*Origin:* Richardson also proposed differentiating between "Client-Side" and "Server-Side" patterns [S2]. With client-side patterns, clients query the Service Registry, select an available instance, and make a request. With server-side pat- terns, clients make requests via a router, which queries the Service Registry and forwards the request to an available instance. However, in the selected works, no implementations reported its usage.

*Goal:* Unlike the API-Gateway pattern, this pattern allows clients and microser- vices to talk to each other directly. It relies on a Service Registry, as depicted in Fig. 2, acting in a similar manner as a DNS server.

*Properties:* The Service Registry knows the dynamic location of each microser- vice instance. When a client requests access to a specific service, it first asks the registry for the service location; the registry contacts the microservice to ensure its availability and forwards the location (usually the IP address or the DNS name and the port) to the calling client. Finally, unlike in the API-Gateway approach, the clients communicate directly with the required services and access all the available APIs exposed by the service, without any filter or service inter- face translation provided by the API-Gateway.

*Evolution and Reported Usage:* A total of eleven papers implemented this pat- tern. Ten of the selected work make a complete usage of the Service Registry style [S13], [S25], [S10], [S9], [S24], [S26], [S30], [S16], and [S38] while [S23] proposes a small variant, implementing the Service Registry by means of an NoSQL database. O'Connor et al. [S36] report on a case study of a partial migration where a legacy SOA system provided some services in connection with new microservices. In this case, the legacy system was accessed like any other microservice. The Service Registry contained the addresses of all microservices and all services provided by the legacy system.

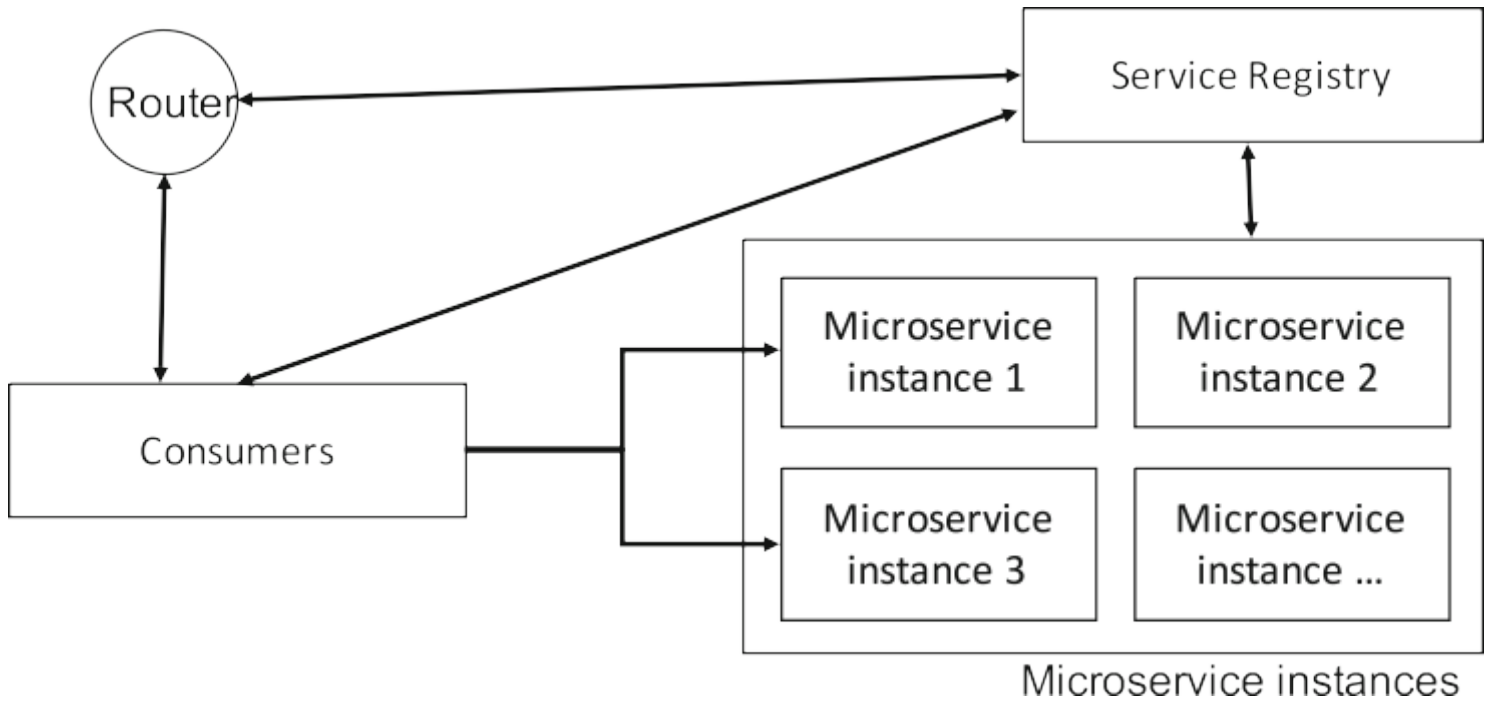

**Fig. 2.** The service registry architectural pattern.

This architectural pattern has several **advantages**:

- *Increased Maintainability*. All the papers reported an increased maintainabil- ity of the systems.
- *Ease of Communication*. Services can communicate with each others directly, without interpretation [S25], [S36].
- *Health Management*. Resilient and scalable mechanisms provide health man- agement and out-scaling functions for atomic and composed services [S7].
- *Failure Safety*. In the case of failure, microservices can be easily restarted, due to their stateless properties [S7].
- *Software Understandability*. Services are relatively small and easy to under- stand [S1], [S2]
- *Ease of Development*. Smaller services are easier to develop [S1], [S2]
- *Ease of Migration*. Existing services can be re-implemented with microser- vices, replacing the legacy service by changing its location in the Service Reg- istry that will start to dynamically serve all microservices instances instead of statically pointing to the legacy system [S36].

Several papers also identified **disadvantages** for this pattern:

- *Interface design must be fixed*. During maintenance, individual services may change internally but there could be a need to also update the interface, requiring adaptation of all connected services. They recommend keeping the interface definition as stable as possible in order to minimize the influence in case of interface changes [S38].
- *Service Registry Complexity*. The registry layer increases implementation complexity as it requires several interfaces per service [S16].

- *Reuse*. If not designed correctly, the service registry could be the main bot- tleneck of the system [S25].
- *Distributed System Complexity*. Direct communication among services increases several aspects: *Communication among Services* [S2], *Distributed Transaction Complexity* [S2], *Testing* of distributed systems, including shared services among different teams can be tricky [S2].

### The Hybrid Pattern

*Concept and Origin:* This pattern combines the power of the Service Registry pattern with that of the API-Gateway pattern, replacing the API-Gateway com- ponent with a message bus.

*Goal and Properties:* Clients communicate with the message bus, which acts as a Service Registry, routing the requests to the requested microservices. Microser- vices communicate with each other through a message bus, in a manner similar to the Enterprise Service Bus used in SOA architectures.

*Evolution and Reported Usage:* Six works implemented this pattern [S27], [S33], [S32], [S35], [S4] and [S3] reporting the following **advantages**:

- *Easy of Migration*. This pattern ease the migration of existing SOA based applications, since the ESB can be used a communication layer for the microservices that gradually replace the legacy services.
- *Learning Curve*. Developers familiar with SOA can easily implement this pattern with a very little training.

and a **disadvantage**:

- *SOA Issues*. The pattern does benefit from the IDEAL properties of microser- vices and from the possibility to independently develop different services with different teams, but has the same ESB-related issues as in SOA.

## 3.3 Deployment Strategies/Patterns

As part of the architectural patterns, we now describe the different deploy- ment strategies (also referred to as deployment patterns) that emerged from our mapping study. Please note that here we only report on microservices-specific deployment strategies not directly related to DevOps, while DevOps automated deployment approaches are reported in Section III.D.

### The Multiple Service per Host Pattern

*Principle:* In this strategy, multiple services and multiple services run on the same host.

*Reported Usage:* Four of the selected works implemented this approach [S19], [S33], [S30], and [S7] without specifying whether they deployed the services into containers or VMs. Fifteen works adopted the same pattern by deploying each service into a container [S10], [S25], [S35], [S9], [S8], [S11], [S34], [S32], [S36], [S38], [S37], [S40], [S16], [S41], and [S22]. Richardson refers to this sub-pattern as "Service instance per container pattern" [S2]. Two works implemented this pattern deploying each microservice into a dedicated virtual machine [S27] and [S31]. This pattern is also called "Service instance per virtual machine" [S2].

Despite reporting on the adoption of these patterns, only a few papers discuss their **advantages** such as:

- *Scalability*. Easy scalability to deploy multiple instances at the same host.
- *Performance*. Multiple containers allow rapid deployment of new services compared to VMs [S40], [S34], [S10].

### The Single Service per Host Pattern

*Principle and Properties:* In this pattern [S2], every service is deployed in its own host. The main benefit of this approach is the complete isolation of ser- vices, reducing the possibility of conflicting resources. However, this dramatically reduces performance and scalability.

*Reported Usage:* This pattern has not been implemented or referenced in the selected works.

## 3.4 Data Storage Patterns

Like any service, microservices need to store data. Sixteen implementations reported on the data storage pattern that they adopted. Among these papers, we identified three different data storage patterns that are also described by [S1], [S24], and [S3]. Although it is recommended to adopt Object Relational Map- ping approaches with NoSQL databases [S1], the patterns identified are also applicable for relational databases.

### The Database per Service Pattern

*Principle and Properties:* In this pattern, each microservice accesses its private database. This is the easiest approach for implementing microservices-based sys- tems, and is often used to migrate existing monoliths with existing databases.

*Reported Usage:* In the selected works, six adopted this pattern [S23], [S12], [S36], [S24], [S11], and [S26]. This pattern has several **advantages**:

- *Scalability*. The database can be easily scaled in a database cluster whithin a second moment [S24], in case the service need to be scaled.
- *Independent Development*. Separate teams can work independently on each service, without affecting other teams in case of changes to the DB schema.
- *Security Mechanism*. Access to other microservices or corruption of data not needed is avoided since only one microservice can access a schema.

### The Database Cluster Pattern

*Principle and Properties:* The second storage pattern proposes storing data on a database cluster. This approach improves the scalability of the system, allowing to move the databases to dedicated hardware. In order to preserve data consis- tency, microservices have a sub-set of database tables that can be accessed only from a single microservice; in other cases, each microservice may have a private database schema. This pattern was described by Richardson [S2].

*Reported Usage:* The pattern was implemented by [S27], [S6], and [S15] by using a separated DB schema for each service. [S15] also proposed it for replicating the data across the DBs of each service.

This pattern has the **advantage** of improving data scalability. It is recom- mended for implementations with huge data traffic while it could be useless in the case of a limited number of users and data traffic. **Disadvantages**:

- *Increased Complexity* through the cluster architecture.
- *Risk of Failure* increases because of the introduction of another component and the distributed mechanism.

### Shared Database Server

*Principle and Properties:* This pattern is similar to the Database Cluster Pat- tern, but, instead of using a database cluster, all microservices access a single shared database.

*Reported Usage:* Six implementations adopted this pattern [S13], [S39], [S25], [S18], [S30], and [S16]. All these implementations access to the data concurrently, without any data isolation approach.

The main **advantage** reported is the simplicity of the migration from mono- lithic applications since existing schemas can be reused without any changes. Moreover, the existing code base can be migrated without the need to make important changes (e.g., the data access layer remains identical).

## 3.5 DevOps and Microservices

Now, we change focus from microservices as an architectural style with principles and patterns to the relevance of the style as a continuous architecting solution. In this section, we answer RQ4, reporting the main DevOps related to tech- niques proposed and applied in conjunction with microservices, summarizing their advantages and disadvantages. The section is structured as follows: After a description of the papers reporting on the application of microservices-based implementations *applying the DevOps pipeline* (partially or completely), we describe the techniques related to each DevOps step: *planning, coding, testing, release, deployment, operation, monitoring.*

**DevOps and Microservices.** Chen et al. [S8] propose a set of tactics for adopting DevOps with microservices-based implementations. Adopting a set of tools enables (1) continuous integration, (2) test automation, (3) rapid deploy- ment and robust operations, (4) synchronized and flexible environment. Their proposal is to keep four quality characteristics under control (availability, modifi- ability, performance, and testability) by means of a set of tactics. As an example, they propose checking availability by monitoring the system by detecting excep- tions, reconfiguring clusters automatically in case of failures or lack of resources, creating active redundancy (by means of Zookeeper [1]) and rolling back deployed services in case of failure.

**Planning and Coding Techniques.** This section includes all techniques and tools for code development, including requirement elicitation, software architec- tures, and coding techniques. As for the architectural styles, we refer to the discussion about the previously described patterns reported in Sects. 3.2 and 3.3. In order to cope with continuously changing requirements whilst ensuring complexity and keeping product evolution under control, Versteden et al. [S16] propose a semantic approach combining microservices and the semantic web. The approach is based on the sharing of a set of ontologies among developers so that they can develop microservices that talk about the same content in the same way. Moreover, this also supports the semantics discoverability of microservices.

Considering the coding activities, Xu et al. [S40] propose "CAOPLE", a new programming language for microservices based on an agent-oriented conceptual model of software systems. This programming language allows defining high- level microservices with the aim of easily developing large connected systems with microservices running in parallel, thus reducing communication overhead among microservices and supporting flexible development.

**Testing Techniques.** (1) Testing is one of the most challenging issues when building a microservice architecture. A microservice architectural style intro- duces several new components into the applications, and more components mean more chances for failure to occur. Automated testing, as one of the main steps of DevOps, has advanced significantly and new tools and techniques continue to be introduced to the market. Therefore, testing a microservices-based application is more complicated because of several issues:

- Because of the language independence of microservices, the testing tools need to be agnostic to any service language or runtime [S42].
- Testing must focus on failure-recovery logic and not on business logic, due to the rapidly evolving code [S42].
- Existing SOA testing methods are commonly not suitable for microservices, since they do not address the two issues aforementioned [S42].

Of the selected papers, seven propose new testing techniques for microservices. However, no implementations report the usage of these techniques.

Testing can be divided into several levels, including unit tests, integration tests, system tests and acceptance tests. As for acceptance tests and system level tests, Rahman et al. [S18] and [S17] propose automating acceptance tests for microservices based on the Behavior-Driven Development (BDD). Since inte- gration, system, and acceptance tests usually need access to the production file system, the network, and the databases, they propose running the tests in the developer's local environment by means of a subset of replicated Docker con- tainers replicating the whole system, including all microservices running in the production environment. They propose running the large test suites in parallel with multiple docker containers deployed on the local development machine so as to allow developers to (1) continue testing on the latest data used in production
(2) continuously run the complete test suites. Unfortunately, they report that running the entire test suite is time consuming and becomes infeasible when the test suite grows. Despite the approach working perfectly for small projects, in big projects the developers' workstations have very high hardware requirements to run the whole system with all microservices and the development environment, making this approach inapplicable in a real development environment.

Savchenko and Radchenko [S22] propose a model of validation of microser- vices that can be executed on local developer machines and in a test environment, before deploy the microservice in production. The model is compliant with the ISO/IEC 29119 [5] and it is based on five steps:

1. Define the interface of every microservice
2. Write unit-tests for each microservice
3. If the unit tests are passed successfully, the microservice can be packed into a container and a set of container self-tests can be executed to ensure that all interfaces defined in the first step are working.
4. If the self-test is passed, then the microservice can be deployed in a test envi- ronment and functional integration tests, load integration tests, and security integration tests can be performed.
5. If all tests in the previous step are passed, the microservice can be deployed in the production environment.

Meinke and Nycander [S20] propose a learning-based testing technique based on a Support Vector Machine (SVM). Thy propose to monitor the inputs of each microservice and to validate the output with a model checker and learn how to interpret the results by means of a SVM based on a stochastic equiva- lence checker. This model is applicable to high-load systems where statistically significant results can be used as training data. It is claimed to be more robust than manual checks since it can test more conditions. However, non-deterministic conditions cannot be verified with this approach, even though they are very rare.

Heorhiadi et al. [S42] propose a network-oriented resiliency testing method to continuously test high-level requirements. They propose a two-level testing platform composed of two layers. The first layer composed by network prox- ies, used to control the communication among microservices, logging any data and reporting communications. The second layer responsible to check the results based on the execution graph, to run the test cases, and, in case of failure, to deploy a new microservice through a "Failure orchestrator". This allows creat- ing and testing long and complex chains of assertions, and validating complex processes. However, the system graph must be provided continuously updated.

Only [S42] has been validated internally by the authors on small sample projects, while the other approaches are only proposals not supported by empir- ical validations. The applicability of the proposed testing techniques to existing large-scale systems therefore needs to be validated.

In conclusion, we can claim that, based on the analysis of the reported test- ing techniques of microservices-based systems, there are no common validation models that support continuous integration of microservices.

**Release Techniques.** No release techniques have been proposed or reported in the selected works.

**Deployment Techniques.** In the selected works, only one work [S31] proposes a technique and a tool for automatic deployment of microservices, assuming the use of reconfigurable microservices. Their tool is based on three main compo- nents: (1) An automatic configuration of distributed systems in OpenStack [4] which, starting from a partial and static description of the target architecture, produces a schema for distributing microservices to different machines and con- tainers; (2) A distributed framework for starting, stopping, and removing ser- vices; and (3) A reconfiguration cordinator which is in charge of interacting the automatic configuration system to produce optimized deployment planning.

**Operation and Monitoring Techniques.** Monitoring cloud services is diffi- cult due to the complexity and distributed nature of the systems. Anwar et al. [S5] highlight the complexity of monitoring task, in particular with microservices- based implementations monitored with OpenStack, reporting that 80% of the commonly collected data are useless, thus collecting only 20% of the actual data would allow analyzing smaller datasets, which are often easier to analyze.

Monitoring is a very important operation at runtime, especially for detect- ing faults in existing services and taking appropriate actions. In this direction, Rajagopalan et al. [S19] propose an autonomous healing mechanism to replace faulty microservices during runtime, in the production environment. They pro- pose comparing the dependency graphs of previous versions of microservices and, in case of failures, replacing the existing microservice by re-deploying the pre- vious version. Despite reducing performance, this approach increases the proba- bility of returning the correct result.

Bak et al. [S21] describe a microservices-based implementation of a dis- tributed IoT case study where they defined an approach for detecting opera- tional anomalies in the system based on the context. They propose an algorithm for detecting records not conformant to the expected or normal behavior of the data, continuously monitoring the various devices and sensors, and dynam- ically building models of typical measurements according to the time of the day. Their anomaly detection approach is based on the analysis of performances and supposed malfunctions. As for failure detection, they also defined a root cause algorithm to understand if some devices crash when located in specific geographic areas because of errors in the data collected from sensors, or crash when connecting to certain devices.

Toffetti et al. [S7] also adopt a self-healing technique in their implementation, simply restarting faulty microservices that return unexpected values or that raise any exceptions, in order to provide the most reliable result.

# 4 Discussion

Most of the implementations reported in the papers are related to research pro- totypes, with the goal of validating the proposed approaches (Table 4). Only six papers report on implementations in industrial context. Regarding the size of the systems implemented, all the implementations are related to small-sized applications, except [S38] that reports on the migration of a large scale system. Only four implementations report on the development language used ([S11], [S32] Java/NodeJS, [S34] php/NodeJS/Python, [S13] php).

## 4.1 Architecture and Deployment Pattern Applications

Several patterns for microservice-based systems emerged from existing imple- mentations (Table 3). We can associate some patterns with specific application settings such as a monolith-to-microservice or SOA-to-microservice migration.

*Migration:* Several implementations report the usage of hybrid systems, aimed at migrating existing SOA-based applications to microservices. Maintenance, and specially independent deployment and the possibility to develop different services with different non-interacting teams, are considered the main reasons for migrating monoliths to microservices. The flexibility to write in different lan- guages and to deploy the services on the most suitable hardware is also consid- ered a very important reason for the migration. Reported migrations from mono- lithic systems tend to be architected with an API-Gateway architecture, proba- bly due to the fact that, since the systems need to be completely re-developed and re-architected, this was done directly with this approach. Migrations from SOA-based systems, on the other hand, tend to have a hybrid pattern, keeping the Enterprise Service Bus as a communication layer between microservices and existing SOA services. Based on this, the Enterprise Service Bus could re-emerge in future evolutions of microservices.

*Deployment:* Another outcome is that deployment of microservices is still not clear. As reported for some implementations, sometimes microservices are deployed in a private virtual machine, requiring complete startup of the whole machine during the deployment, thus defeating the possibility of quick deploy- ment and decreasing system maintainability due to the need for maintaining a dedicated operating system, service container, and all VM-related tasks.

**Table 3.** Classification of advantages and disadvantages of the identified patterns [16].

| | Pattern | Advantages | Disadvantages |
|---|---|---|---|
| Orchestration & coordination | General | - Increased maintainability<br>- Can use different languages<br>- Flexibility<br>- Reuse<br>- Physical isolation<br>- Self-healing | - Development, Testing, Complexity<br>- Implementation effort<br>- Network-related issue |
| | API gateway | - Extension easiness<br>- Market-centric architecture<br>- Backward compatibility | - Potential bottleneck<br>- Development complexity<br>- Scalability |
| | Service registry | - Increased maintainability<br>- Communic., developm., migration<br>- Software understandability<br>- Failure safety | -Interface design must be fixed<br>- Service registry complexity<br>- Reuse<br>- Distributed system complexity |
| | Hybrid | - Migration easiness<br>- Learning curve | - SOA/ESB integration issues |
| Deploy | Multiple service per host | - Scalability<br>- Performance | |
| | Single service per host | - Service isolation | - Scalability<br>- Performance |
| Data storage | DB per service | - Scalability<br>- Independent development<br>- Security mechanism | - Data needs to be splitted<br>- Data consistency |
| | DB cluster | - Scalability<br>- Implementation easiness | - Increase complexity<br>- Failure risks |
| | Shared DB server | - Migration easiness<br>- Data consistency | - Lack of data isolation<br>- Scalability |

**Table 4.** The implementations reported in the selected works [16].

| | Research prototype | Validation-specific implementations | Industrial implementations |
|---|---|---|---|
| Websites | - [S11], [S39] | - [S15], [S24], [S26], [S31] | - [S13], [S32] |
| Services/API | - IOT integration [S33] | - [S9], [S10], [S14], [S16], [S23], [S37], [S36] | - [S21], [S34] |
| Others | - Enterprise measurement system [S4]<br>- IP multimedia system [S25] | - Benchmark/Test [S35], [S41], [S42]<br>- Business process modelling [S12] | - Mobile dev. platform [S38]<br>- Deployment platform [S30] |

## 4.2 DevOps Link

Taking into account the continuous delivery process, the DevOps pipeline is only partially covered by research work. Considering the idea of continuous architect- ing, there is a number of implementations that report success stories regarding how to architect, build, and code microservice-based systems, but there are no reports on how to continuously deliver and how to continuously re-architect exist- ing systems. As reported in our classification schema of the research on DevOps techniques (Table 5), the operation side, monitoring, deployment, and testing techniques are the most investigated steps of the DevOps pipeline. However, only few papers propose specific techniques, and apply them to small example projects. Release-specific techniques have not been investigated in our selected works. No empirical validation have been carried out in the selected works. Therefore, we believe this could be an interesting result for practitioners, to understand how existing testing techniques can adopted in industry.

**Table 5.** DevOps techniques classification schema.

| | Proposed technique |
|---|---|
| Planning | - Semantic models [S16] |
| Coding | - Agent-oriented programming language [S40] |
| Testing | - BDD automated acceptance test [S17], [S18]<br>- SVM learning-based testing [S20]<br>- Validation on developers' machine [S22]<br>- Resiliency test of high-level requirements [S42] |
| Release | |
| Deployment | - Automated deployment [S31] |
| Monitoring | - Self-healing to replace faulty MS [S7], [S19] |
| Operation | - Context-based anomalies detection [S21] |

## 4.3 Research Trends and Gaps

*Industry First:* Different research trends have emerged in this study. First of all, we can see that microservices come from practitioners and research comes later, so reports on existing practices are only published with delay. From the architec- tural point of view, the trend is to first analyze the industrial implementations and then compare them with previous solutions (monolithics or SOA).

*Style Variants:* A new microservice architectural styles variant was proposed by researchers ([S24] and [S26]), applying a database approach for microservice orchestration. However, because they have just been published, no implemen- tations have adopted these practices yet. Also in this case, we believe that an empirical validation and a set of benchmarks comparing this new style with existing one could be highly beneficial for researchers and practitioners.

Despite the increasing popularity of microservices and DevOps in industry, this work shows the lack of empirical studies in an industrial context reporting how practitioners are continuously delivering value in existing large-scale sys- tems. We believe that a set of studies on the operational side of the DevOps pipeline could be highly beneficial for practitioners and could help researchers understand how to improve the continuous delivery of new services.

We can compile the following **research gaps** and **emerging issues**:

- *Position Papers and Introduction to microservices.* An interesting outcome of this work, obtained thru the reading of the whole literature is the tendency of publishing several position papers, highlighting some microservices properties or reporting about potential issues, without any empirical evidence.
- *Comparison of SOA and Microservices.* The differences have not been thoroughly investigated. There is a lack of comparison from different points of view (e.g., performance, development effort, maintenance).
- *Microservices Explosion.* What happens once a growing system has thousands/millions of microservices? Will all aforementioned qualities degrade?
- *DevOps related techniques.* Which chain of tools and techniques is most suitable for different contexts?
- *Negative Results.* In which contexts do microservices turn out to be counterproductive? Are there anti-patterns [6,15,17]?

### 4.4 Towards an Integrated Microservice Architecture and Deployment Perspective

Further to the discussion of trends and gaps that we have provided in the previ- ous subsection, we focus a short discussion here on an aspect that has emerged from the discussion of DevOps and Microservices in the section before. Automa- tion and tool support are critical concerns for the deployment of microservices for instance in the form of containers, but also the wider implementation of microservice architectures in a DevOps pipeline with tool support for continu- ous integration and deployment.

In [7], the success of mmicroservices is linked to the evolution of technology platforms.

- Containerization with LXC or Docker has been the first wave, enabling the independent deployment of microservices.
- Container orchestration based on Mesos, Kubernestes or Docker Swarm enables better management of microservices in distributed environments.
- Continuous delivery platform such as Ansible or Spinnaker have also had its impact as our DevOps discussion shows.

Currently, further technologies are finding their way into architecting soft- ware:

- Serverless computing fociussing on function-as-a-service solutions that allow more fine-grained service functions without the need to be concerned with infrastructure resources.
- Service meshes address the need fully integrated service-to-service communi- cation monitoring and management.

This indicates that as the technology landscape evolves, we can expect new patterns to emerge. Thus pattern identification will remain a task for the future.

## 5 Threats to Validity

Different types of threats to validity need to be addressed in this study.

*Construct validity* reflects what is investigated according to the research ques- tions. The terms microservices, DevOps, and all sub-terms identified in Table I are sufficiently stable to be used as search strings. In order to assure the retrieval of all papers on the selected topic, we searched broadly in general publication databases, which index most well-reputed publications. Moreover, we included gray literature if their citations were higher than the average, in order to con- sider relevant opinions reported in non-scientific papers. Reliability focuses on whether the data are collected and the analysis is conducted in a way that can be repeated by other researchers with the same results. We defined search terms and applied procedures that can be replicated by others. Since this is a map- ping study and not a systematic review, the inclusion/exclusion criteria are only related to whether the topic of microservices is present in a paper [9].

*Internal validity* is concerned with data analysis. Since our analysis only uses descriptive statistics, the threats are minimal.

*External validity* is about generalization from this study. Since we do not draw any conclusions about mapping studies in general, external validity threats are not applicable.

## 6 Conclusion

In this work, we conducted a systematic mapping study on micro-services-based architectural style principles and patterns, also looking at techniques and tools for continuously delivering new services by applying the DevOps approach when implementing micro-services-based systems.

As main outcome we identified several research gaps, such as the lack of comparison between SOA and Microservices, the investigation of consequences of microservices explosion and the high interest in exploring microservices in DevOps settings. Most of the selected works were published at workshops or conferences, which confirms the novelty of this topic and the interest in con- ducting this mapping study.

We have used architectural patterns to identify common structural prop- erties of microservice architectures. Three orchestration and data-storage pat- terns emerged that appear to be widely applied for microservices-based sys- tems. Although some patterns were clearly used for migrating existing mono- lithic applications (service registry pattern) and others for migrating existing SOA applications (hybrid pattern), adopting the API-Gateway pattern in the orchestration layer in order to benefit from microservice architectures without refactoring a second time emerges as a key recommendation. Overall, a 3-layered catalog of patterns comes out with patterns for orchestration/coordination and storage as structural patterns and for deployment alternatives.

Independent deployability, being based on strong isolation, and easing the deployment and self-management activities such as scaling and self-healing, and also maintainability and reuse as classical architecture concerns are the most widely agreed beneficial principles.

DevOps in the contest of microservices is an hot topic being frequently dis- cussed online among practitioners, despite small number of works, probably because of its novelty. Work in this topic is mainly covering testing and monitor- ing techniques, while there are not yet papers on release techniques. Nonetheless, the independent deployability property often cited requires microservices to be mapped to a continuous architecting pipeline. Therefore, we believe DevOps would need more empirical validation in the context of microservices.

A further analysis regards the notion of a architecture style itself in case of continuous architecting. The latter becomes an integral element of software architecture these days. Correspondingly, an architectural style requires to cover continuous architecting activities as well in addition to purely development stage regards such as system design usually focused on in architectural styles.

## A The Selected Studies

[S1] Lewis, J. and Fowler, M. 2014. Microservices. http://martinfowler.com/articles/microservices.html.

[S2] Richardson, C. 2014. Microservice Architecture http://microservices.io.

[S3] Namiot, D. and Sneps-Sneppe, M. 2014. On micro-services architecture. Inter- national Journal of Open Information Technologies V.2(9).

[S4] Vianden, M., Lichter, H. and Steffens, A. 2014. Experience on a Microservice- Based Reference Architecture for Measurement Systems. Asia-Pacific Software Engineering Conference.

[S5] Anwar, A., Sailer, A., Kochut, A., Butt, A. 2015. Anatomy of Cloud Monitoring and Metering: A Case Study and Open Problems. Asia-Pacific Workshop on Systems.

[S6] Patanjali, S., Truninger, B., Harsh, P. and Bohnert, T. M. 2015. CYCLOPS: A micro service based approach for dynamic rating, charging & billing for cloud. Conference on Telecommunications.

[S7] Toffetti, G., Brunner, S., Blöchlinger, M., Dudouet, F. and Edmonds. A. 2015. Architecture for Self-managing Microservices. Int. Workshop on Automated Inci- dent Mgmt in Cloud.

[S8] Chen, H.M., Kazman, R., Haziyev, S.,Kropov, V. and Chtchourov, D. 2015. Architectural Support for DevOps in a Neo-Metropolis BDaaS Platform. Symp. on Reliable Distr Syst Workshop.

[S9] Stubbs, J., Moreira, W. and Dooley, R. 2015. Distributed Systems of Microser- vices Using Docker and Serfnode. Int. Workshop on Science Gateways.

[S10] Abdelbaky, M., Diaz-Montes, J., Parashar, M., Unuvar, M. and Steinder, M. 2015. Docker containers across multiple clouds and data center. Utility and Cloud Computing Conference.

[S11] Villamizar, M., Garcas, O., Castro, H. et al. 2015. Evaluating the monolithic and the microservice architecture pattern to deploy web applications in the cloud. Computing Colombian Conference

[S12] Alpers, S., Becker, C., Oberweis, A. and Schuster, T. 2015. Microservice Based Tool Support for Business Process Modelling. Enterprise Distributed Object Computing Workshop.

[S13] Le, V.D., Neff, M.M., Stewart,R.V., Kelley, R., Fritzinger, E., Dascalu, S.M. and Harris, F.C. 2015. Microservice-based architecture for the NRDC. Industrial Informatics Conference.

[S14] Malavalli, D. and Sathappan, S. 2015. Scalable Microservice Based Architecture for Enabling DMTF Profiles. Int. Conf. on Network and Service Management.

[S15] Viennot, N., Mathias, M. Lécuyer, Bell, J., Geambasu, R. and Nieh, J. 2015. Synapse: A Microservices Architecture for Heterogeneous-database Web Appli- cations. European Conf. on Computer Systems.

[S16] Versteden, A., Pauwels, E. and Papantoniou, A. 2015. An ecosystem of user- facing microservices supported by semantic models. International USEWOD Workshop.

[S17] Rahman, M. and Gao, J. 2015. A Reusable Automated Acceptance Testing Archi- tecture for Microservices in Behavior-Driven Development. Service-Oriented Sys- tem Engineering Symp.

[S18] Rahman, M., Chen, Z. and Gao, J. 2015. A Service Framework for Parallel Test Execution on a Developer's Local Development Workstation. Service-Oriented System Engineering Symp.

[S19] Rajagopalan, S. and Jamjoom, H. 2015. App-Bisect: Autonomous Healing for Microservice-based Apps. USENIX Conference on Hot Topics in Cloud Comput- ing.

[S20] Meink, K. and Nycander, P. 2015. Learning-based testing of distributed microser- vice architectures: Correctness and fault injection". Software Engineering and Formal Methods workshop.

[S21] Bak, P., Melamed, R., Moshkovich, D., Nardi, Y., Ship, H., Yaeli, A. 2015. Loca- tion and Context-Based Microservices for Mobile and Internet of Things Work- loads. Conference Mobile Services.

[S22] Savchenko, D. and Rodchenko, G. 2015. Microservices validation: Methodology and implementation. Ural Workshop on Parallel, Distributed, and Cloud Com- puting for Young Scientists.

[S23] Gadea, C., Trifan, M., Ionescu, D., Ionescu, B. 2016. A Reference Architecture for Real-time Microservice API Consumption. Workshop on CrossCloud Infras- tructures & Platforms.

[S24] Messina, A., Rizzo, R., Storniolo, P., Urso, A. 2016. A Simplified Database Pat- tern for the Microservice Architecture. Adv. in Databases, Knowledge, and Data Applications.

[S25] Potvin, P., Nabaee, M., Labeau, F., Nguyen, K. and Cheriet, M. 2016. Micro service cloud computing pattern for next generation networks. EAI International Summit.

[S26] Messina, A., Rizzo, R., Storniolo, P., Tripiciano, M. and Urso, A. 2016. The database-is-the-service pattern for microservice architectures. Information Tech- nologies in Bio-and Medical Informatics conference.

[S27] Leymann, F., Fehling, C., Wagner, S., Wettinger, J. 2016. Native cloud applica- tions why virtual machines, images and containers miss the point. Cloud Comp and Service Science conference.

[S28] Killalea, T. 2016. The Hidden Dividends of Microservices. Communications of the ACM. V.59(8), pp. 42-45.

[S29] M. Gysel, L. Kölbener, W. Giersche, O. Zimmermann. "Service cutter: A sys- tematic approach to service decomposition". European Confeence on Service- Oriented and Cloud Computing.

[S30] Guo, D., Wang, W., Zeng,G. and Wei, Z. 2016. Microservices architecture based cloudware deployment platform for service computing. Symposyum on Service- Oriented System Engineering. 2016.
[S31] Gabbrielli, M., Giallorenzo, S., Guidi, C., Mauro, J. and Montesi, F. 2016. Self-Reconfiguring Microservices". Theory and Practice of Formal Methods.
[S32] Gadea, M., Trifan, D. Ionescu, et al. 2016. A microservices architecture for col- laborative document editing enhanced with face recognition. SAC.
[S33] Vresk, T. and Cavrak, I. 2016. Architecture of an interoperable IoT platform based on microservices. Information and Communication Technology, Electronics and Microelectronics Conference.
[S34] Scarborough, W., Arnold, C. and Dahan, M. 2016. Case Study: Microservice Evolution and Software Lifecycle of the XSEDE User Portal API. Conference on Diversity, Big Data & Science at Scale.
[S35] Kewley, R., Keste, N. and McDonnell, J. 2016. DEVS Distributed Modeling Framework: A Parallel DEVS Implementation via Microservices. Symposium on Theory of Modeling & Simulation.
[S36] O'Connor, R., Elger, P., Clarke, P., Paul, M. 2016. Exploring the impact of situ- ational context - A case study of a software development process for a microser- vices architecture. International Conference on Software and System Processes.
[S37] Jaramillo, D., Nguyen, D. V. and Smart, R. 2016. Leveraging microservices archi- tecture by using Docker technology SoutheastCon.
[S38] Balalaie, A., Heydarnoori, A. and Jamshidi, P. 2015. Migrating to Cloud-Native architectures using microservices: An experience report. European Conference on Service-Oriented and Cloud Computing.
[S39] Lin, J. Lin, L.C. and Huang, S. 2016. Migrating web applications to clouds with microservice architectures. Conference on Applied System Innovation.
[S40] Xu,C., Zhu, H., Bayley, I., Lightfoot, D., Green, M. and Marshall P. 2016. CAOPLE: A programming language for microservices SaaS. Symp. on Service- Oriented System Engineering.
[S41] Amaral, M., Polo, J., Carrera, D., et al. 2015. Performance evaluation of microser- vices architectures using containers. Int. Symp. on Network Computing and Applications.
[S42] Heorhiadi, V., Rajagopalan, S., Jamjoom, H., Reiter, M.K., and Sekar, V. 2016. Gremlin: Systematic Resilience Testing of Microservices. International Confer- ence on Distributed Computing Systems.

## References

1. Apache ZooKeeper: https://zookeeper.apache.org/
2. Balalaie, A., Heydarnoori, A., Jamshidi, P.: Microservices architecture enables DevOps: migration to a cloud-native architecture. IEEE Softw. **33**(3), 42–52 (2016)
3. Bass, L., Weber, I., Zhu, L.: DevOps: A Software Architects Perspective, 1st edn. Addison-Wesley Professional, Boston (2015)
4. Di Cosmo, R., Eiche, A., Mauro, J., Zacchiroli, S., Zavattaro, G., Zwolakowski, J.: Automatic deployment of services in the cloud with aeolus blender. In: Barros, A., Grigori, D., Narendra, N.C., Dam, H.K. (eds.) ICSOC 2015. LNCS, vol. 9435, pp. 397–411. Springer, Heidelberg (2015). https://doi.org/10.1007/978-3-662-48616- 0 28
5. ISO/IEC/IEEE 29119 Software Testing (2014). http://www. softwaretestingstandard.org/

6. Jamshidi, P., Pahl, C., Mendonca, N.C.: Pattern-based multi-cloud architecture migration. Softw. Pract. Exp. **47**(9), 1159–1184 (2017)
7. Jamshidi, P., Pahl, C., Mendonca, N.C., Lewis, J., Tilkov, S.: Microservices: the journey so far and challenges ahead. IEEE Softw. **35**(3), 24–35 (2018)
8. Kitchenham, B., Charters, S.: Guidelines for Performing Systematic Literature Reviews in Software Engineering (2007)
9. Kitchenham, B., Brereton, P.: A systematic review of systematic review process research in software engineering. Inf. Softw. Technol. **55**(12), 2049–2075 (2013)
10. Lewis, J., Fowler, M.: MicroServices (2014). www.martinfowler.com/articles/microservices.html
11. Pahl, C., Jamshidi, P.: Microservices: a systematic mapping study. In: International Conference on Cloud Computing and Services Science (2016)
12. Pahl, C., Jamshidi, P., Zimmermann, O.: Architectural principles for cloud soft- ware. ACM Trans. Internet Technol. **18**(2), 17 (2018)
13. Petersen, K., Feldt, R., Mujtaba, S., Mattsson, M.: Systematic mapping studies in software engineering. In: EASE (2008)
14. Richardson, C.: Decomposing Applications for Deployability and Scalability, Microservices (2014). https://www.infoq.com/articles/microservices-intro
15. Taibi, D., Lenarduzzi, V., Pahl, C.: Processes, motivations, and issues for migrating to microservices architectures: an empirical investigation. IEEE Cloud Comput. **4**(5), 22–32 (2017)
16. Taibi, D., Lenarduzzi, V., Pahl, C.: Architectural patterns for microservices: a systematic mapping study. In: International Conference on Cloud Computing and Services Science, pp. 221–232 (2018)
17. Taibi, D., Lenarduzzi, V.: On the definition of microservices bad architectural smells. IEEE Softw. **35**(3), 56–62 (2018)
18. Taibi, D., Systä, K.: From monolithic systems to microservices: a decomposition framework based on process mining. In: 8th International Conference on Cloud Computing and Services Science, CLOSER (2019)
19. Wohlin, C.: Guidelines for snowballing in systematic literature studies and a repli- cation in software engineering. In: EASE 2014 (2014)